\documentclass[preprint,aps,prb,showpacs,amsmath,amssymb]{revtex4}
\usepackage{graphics}
\usepackage{epsf,graphicx,pstcol,epsfig,psfrag}
\begin{document}

\title{Thermodynamics of weakly coupled Falicov-Kimball chains from renormalization-group theory}
\author{Jozef Sznajd}
\affiliation{Institute for Low Temperature and Structure Research, Polish Academy of Sciences, Wroclaw} 
\date{\today}

\begin{abstract}

The linear perturbation renormalization group is used to study spinless two-band fermion chains at half-filling. The model consists of two species of spinless fermions, localized $f$ and extended $p$ and takes into account: the kinetic energy of fermions $p$, the on-site Coulomb repulsion $V$ between $p$ and $f$ fermions, chemical potentials $\mu_p$ and $\mu_f$ adjusted in such a way that the average of the site occupation $<n^i_f> + <n^i_p> = 1$ and a weak interchain hopping $t_x$. The average occupation number, the specific heat and the correlation functions are studied as functions of temperature. For a single chain the occupation number is a smooth function of T and the specific heat displays two maxima. The weak interchain hopping triggers a discontinuity in the occupation number of fermions as a function of temperature. A long-standing controversy on whether the Falicov-Kimball model can describe a discontinuous transition of $n_f$ is also addressed.

\end{abstract}


\maketitle
\section{The model}
\label{introduction}
Spinless fermions \cite{Kohn} can be considered as fully polarized electrons in a high magnetic field, but usually they are studied as a simplified model for the spin-$\frac{1}{2}$ fermions. Such a model is, of course, magnetically uninteresting, but it can still have interesting features associated with the effect of a competition between Coulomb repulsion and kinetic energy as well as interband mixing or charge ordering. For the single-band spinless model at half filling and zero temperature it is expected that a growth of the Coulomb repulsion leads to a transition from a metallic to an insulating charge-ordered state. The evaluation of physical quantities at finite temperature possess some difficulties even for one-dimensional integrable models. Thermodynamics of such a model has been discussed by the thermodynamic Bethe ansatz and by the quantum transfer matrix approach \cite{Sakai}. In our previous paper \cite{SB} we have used to study the weakly coupled chains of the spinless model the linear renormalization group transformation and found the metal-insulator phase transition temperature as a function of interchain hopping parameter. 

A single-band spinless model is not sufficient to describe all the relevant physics of the highly correlated electron systems. Therefore, to describe strong electron-electron correlations several extensions and generalizations  of the single-band spinless model have been proposed. And so, the spinless periodic Anderson model with phonons \cite{Grewe} and the extended Hubbard model with spinless itinerant and localized electrons \cite{Sch, HH, Barma} were used in the past for the description of the mixed-valance systems. The latter model in the limit of infinite dimensions was studied in the context of the metal-insulator transition \cite{Si}. The renormalization-group equation was derived for the two-band spinless fermion model in one dimension by Muttalib and Emery \cite{Emery}.
In this paper we study the thermodynamic properties of the simplified two-band Hubbard model without hybridization, proposed by Falicov and Kimball \cite{FK}  to describe the metal-semiconductor transitions in metallic oxides. 
The Falicov-Kimball model (FKM) is one of the simplest nontrivial interacting electron models and since its creation it has attracted much attention in the literature. It has been used to study several phenomena such as metal-insulator transitions \cite{RFK, MP}, phase separation in the binary alloy \cite{FGM}, intermediate valence \cite{JWS} and charge density wave order \cite{FL} to mention the most common examples.

Except for very few rigorously controlled results in the strong-coupling regime and low temperature \cite{Datta}, much of the finite temperature results for the FKM have either been based  on the molecular field approximation or they are restricted to one \cite{PF, MAS} or  an infinite number of spatial dimensions \cite{JF}. Recently the Suzuki-Takano renormalization group  transformation \cite{Suz} combined with the Migdal-Kadanoff bond moving approximation \cite{MK} was used to find the phase diagram of the FKM \cite{Berker}. The authors obtained the global phase diagram  of the $d=3$ FKM for whole range of interactions (hopping, on-site Coulomb repulsion and chemical potentials). However, they have not studied the temperature dependence of the thermodynamic quantities and we should notice that the Migdal-Kadanoff approximation badly reproduces the physical content of the simplest $s=\frac{1}{2}$ field-free Heisenberg model and gives rather poor quantitative results even for the 2d Ising model.

The model considered in this paper is made of an infinite number of spinless fermion chains coupled by a weak interchain hopping. The model consists of two species of spinless fermions: localized, denoted by $f$ and extended $p$ and can be defined by the following Hamiltonian:

\begin{equation}
H = H_0+H_I,
\end{equation}
where $H_0$ denotes one chain Hamiltonian,
\begin{eqnarray}
\label{1} 
H_0 &=&\tilde{t_p}\sum_{\big < ij\big>}( p_i^\dagger p_j +
p_j^\dagger p_i)+\tilde{V} \sum_i n^i_f n^i_p +\tilde{\mu_p} \sum_i n^i_p+\tilde{\mu_f} \sum_i n^i_f,
\end{eqnarray}
and
\begin{equation}
\tilde{t_p}=\frac{t_p}{T}, \quad \tilde{V}=\frac{V}{T}, \quad \tilde{\mu_i} = \frac{\mu_i}{T},
\end{equation}
$p_i^\dagger$, $p_i$ $(n^i_p=p_i^\dagger p_i)$ are the creation and annihilation operators of the itinerant spinless fermions and $ f_i^\dagger$, $f_i$ $(n^i_f=f_i^\dagger f_i)$  the creation and annihilation operators for the spinless fermions in the localized state.  \emph {A factor $-\beta = -1/k_B T$ has been absorbed in the Hamiltonian (1)}. Thus,  $V < 0$ means the repulsive Coulomb interaction. The first term in (2) is the kinetic energy corresponding to the hopping of the itinerant  fermions $p$ between sites $ i$ and $j$.  The second term represents the on-site Coulomb repulsion between $p$ itinerant and $f$ localized fermions, $\mu_p$ and $\mu_f$ are chemical potentials adjusted in such a way that the average of the site occupation 
\begin{equation}
<n^i_f>+<n^i_p> = 1.
\end{equation}
In this paper we assume that the localized level $\mu_f$ is temperature independent and the condition (3) is fulfilled by calculating the chemical potential $\mu_p$.
For $t_p=0$ the model (1) describes two classical subsystems (Ising models),  both chemical potentials are temperature independent and the condition $<n^i_f>+<n^i_p> = 1$ is fulfilled for
\begin{equation}
\mu_f + \mu_p = -V.
\end{equation}
In the symmetric case $\mu_f = \mu_p = -V/2$, the average occupation numbers  (AON)  $<n_f> = <n_p> = 1/2$ whereas for $\mu_f \neq \mu_p$ the AON are temperature dependent. In Fig.1 the average occupation numbers for $t_p=0$ and two cases:  (i) $\mu_f = 0,  \mu_p =-V$ and (ii)  $\mu_f = 2.1,  \mu_p =1.9, V = -4$ are presented.  

\begin{figure}
\label{Fig_1}
 \epsfxsize=10cm \epsfbox{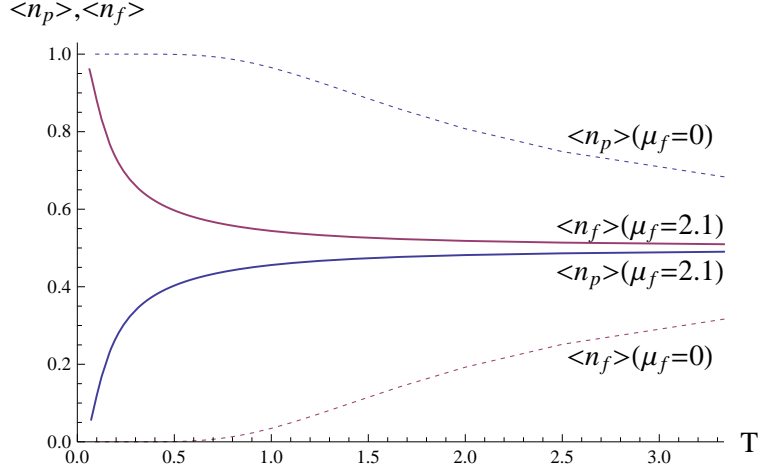}
 \caption{ (Color online) Averages of the site occupations $<n^f_i>$ and $<n^p_i>$ as functions of temperature for $t_p = 0$, $V = -4$ and $\mu_f =0$ (dashed lines) and $\mu_f = 2.1$ (solid lines). }
 \end{figure}

The interchain coupling is restricted to the hopping of the $p$ fermions between the nearest neighbor chains and $H_I$ reads

\begin{equation}
H_I = \tilde{t_x} \sum_{ i,n}( p_{i,n}^\dagger p_{i,n+1} +
p_{i,n+1}^\dagger p_{i,n}) 
\end{equation}
where $n$ numbers the chains.  In the following all values will be given in units of $t_p$ ($t_p=1$).

The purpose of this paper is to study by using the linear perturbation renormalization group, the thermodynamics: the specific heat, correlation functions, and average occupation number of the $p$ fermions, as functions of temperature of the two-band spinless fermion model (FKM). The discontinuous transition of the $p$ ($f$)-fermion occupation number $<n_p>$ ($<n_f>$) as function of temperature is also discussed.

\section{Linear Perturbation Renormalization Group}

The Linear Perturbation Renormalization Group (LPRG) approach starts with an approximate decimation of one chain (2). Then, on the basis of it the interchain interaction is renormalized in a perturbative way \cite{JS}. 

The RG transformation for the Hamiltonian (1) is defined by

\begin{equation}
e^{{ H'}({\hat \sigma,\hat \phi})} = Tr_p P({\hat \sigma,\hat \phi};{\hat p, \hat f}) e^{{H}({\hat p,\hat f})}
\end{equation}
with a linear weight operator $P({\hat \sigma,\hat \phi};{\hat p, \hat f})$ which projects the original fermions $p$ and $f$ space onto the space of the new fermions $\sigma$ and $\phi$, 
\begin{eqnarray}
P({\hat \sigma,\hat \phi};{\hat p, \hat f}) = &\prod_{i=0} & (1+p_{mi+1}^\dagger \sigma_{i+1}+\sigma_{i+1}^\dagger p_{mi+1} + 2 n^{mi+1}_{p} n^{i+1}_{\sigma} -n^{mi+1}_{p}-n^{i+1}_{\sigma}) \nonumber\\
&\times& (1+f_{mi+1}^\dagger \phi_{i+1}+\phi_{i+1}^\dagger f_{mi+1} + 2 n^{mi+1}_{f} n^{i+1}_{\phi} -n^{mi+1}_{f}-n^{i+1}_{\phi}),
\end{eqnarray}
where $\sigma_i^\dagger, \sigma_i$ $(n^i_{\sigma}=\sigma_i^\dagger \sigma_i)$ and $\phi_i^\dagger, \phi_i$ $(n^i_{\phi}=\phi_i^\dagger \phi_i)$ are the creation and annihilation operators of the new spinless fermions. 
For a single chain (2) the transformation (6,7) is a Suzuki-Takano \cite{Suz}-type decimation transformation. For instance, for $m = 2$ in each renormalization step every other site survives, whereas for $m = 3$ every third site survives, and so on.  In order to obtain effective interactions between the operators on surviving sites, for $m = 2$ a three site block has to be considered, and generally a $m+1$ - site block for any $m$. 
     
For an infinite number of chains the RG transformation (5) can be written as

\begin{equation}
H' = \ln Tr_{p,f}  P e^{ H_0 + H_I}.
\end{equation}
For simplicity, from now we omit arguments of the operators $H$ and $P$. Because of noncommutativity of several parts of the Hamiltonian there is a necessity of an approximate decomposition of the exponential operator. The simplest second order decomposition is given by the symmetric product \cite{Suz1} 
\begin{equation}
e^{ H_0+ H_I}  \approx  e^{\frac{H_ 0 } {2} }  e^{ H_ I}  e^{\frac{H_ 0 } {2} }
\end{equation}
The interchain interaction ${\cal H}_I$ is renormalized in a perturbative way and
if we confine ourselves to the second order in the cumulant expansion the transformation (8) can be rewritten as
\begin{eqnarray}
 H' & =& \ln z_0+\frac{1}{2}( <H_ I>_P+ <H_ I>_L) \nonumber \\
&+& \frac{1}{8}(< H_ I^2>_P+2< H_ I^2>_{P-L}+<H_ I^2>_L) \nonumber \\ &-& \frac{1}{8}(<H_ I>_P^2+2<H_ I>_P< H_ I>_L+< H_ I>_L^2),
\end{eqnarray}
where 
\begin{eqnarray}
z_0 = Tr_{p,f} P e^{ H_0}, \qquad \qquad \qquad  <A>_L =\frac{1}{z_0} Tr_{p,f} P A e^{ H_0}, \nonumber \\ <A>_P =\frac{1}{z_0} Tr_{p,f} P  e^{H_0 A}, \quad 
<A^2>_{P-L} =\frac{1}{z_0} Tr_{p,f} P A e^{H_0} A.
\end{eqnarray}

In contrast to the one band model \cite{SB} in the case of the Hamiltonian (2) the RG transformation generates new interactions even for a single chain. So, except for the four original parameters (2) - $t_p, V, \mu_p$ and $\mu_f$ in the renormalization procedure the following eight new couplings come into play:
\begin{eqnarray}
&&u_p n^i_p n^j_p,\quad u_f n^i_f n^j_f,\quad v n^i_p n^j_f,\quad g_1 n^i_p n^j_p n^i_f,\quad g_2 n^i_f n^j_f n^i_p,\nonumber\\
&&\quad g_4 n^i_f n^j_f n^i_p n^j_p,\quad g_p p^\dagger_i p_j n^i_f,\quad g_n p^\dagger_i p_j n^i_f n^j_f.
\end{eqnarray}

If one considers the chains in higher dimensions the LPRG transformation (10) generates additional interactions.
The number of these new interactions already in the lowest nontrivial order cumulant expansion is infinite for an infinite system \cite{JS, JS1}. So, the LPRG transformation is obtained by using several approximations: the abbreviation of the cumulant expansion (10), the truncation of the new interchain interactions generated by the transformation, the approximate decomposition of the exponential operator (9), and the block approximation used for one-dimensional decimation \cite{Suz}. All of these approximations are high-temperature approximation. Thus, the LPRG is an approach reliable at rather high temperatures.

 \section{Two band spinless fermion chain}

As mentioned above applying the transformation (6) with the projector (7)  for any $m$ to the single chain Hamiltonian (2) one obtains renormalized Hamiltonian ${\cal H'}$ for new fermion operators $(\sigma^\dagger, \sigma,\phi^\dagger, \phi)$ and renormalized parameters $t'_p, \mu'_p, \mu'_f$ and $V'$ with additional eight terms (12) generated by the transformation. So, one has to complete the original Hamiltonian (2) by those couplings, and finally considers the renormalization group flow in twelve dimensional coupling parameters space. For $m=3$ (four site block) the renormalized couplings read:

\begin{eqnarray}
t'_p = \frac{1}{2} \lg \frac{\lambda_5}{\lambda_4},\quad \mu'_f = 2 \lg \frac{\lambda_2}{f_0}, 
\quad \mu'_p = \lg \frac{\lambda_4 \lambda_5}{f_0^2}, \quad
V' = \lg \frac {f_0^2 \lambda_6  \lambda_8} {\lambda_2^2 \lambda_4  \lambda_5} - \frac {f_{V}-f_{v}} {W}   \log\frac{\lambda_6}{\lambda_8},     
\end{eqnarray}
where
\begin{eqnarray}
W =\sqrt{4 f_{g_p}^2+8 f_{g_p} f_{t_p}+4 f_{t_p}^2+(f_{V}-f_{v})^2}, 
\end{eqnarray}
and $\lambda_i$ are eigenvalues of the operator $H_0$ 
\begin{eqnarray}
\lambda_2=f_0+f_{\mu_f},\quad  \lambda_{4,5} =f_0+f_{\mu_p} \mp f_{t_p}, \quad  
\lambda_{6,8} = \frac{1}{2}(2 f_0+2f_{\mu_p}+2f_{\mu_f}  +f_{V}+f_{v} \mp W),
\end{eqnarray}
with
\begin{eqnarray} 
\label{12}
&& f_0=Tr_{p,f} R_{f} R_{p} e^{H_0}, \quad f_t=Tr_{p,f} p_1 p_4^{\dagger} e^{ H_0}, \nonumber \\
&& f_{\mu_p}=Tr_{p,f} (2 n_{1p}+n_{4p}-2 n_{1p} n_{4p}-1) R_{f} e^{H_0}, 
\nonumber \\
&& f_{\mu_f}=Tr_{p,f} (2 n_{1f}+n_{4f}-2 n_{1f} n_{4f}-1) R_{p} e^{H_0}, 
\nonumber \\
&& f_{V}=Tr_{p,f} (2 n_{1p}-1)(1-n_{4p})(2 n_{1f}-1)(1-n_{4f}) e^{H_0}.
\end{eqnarray}
and
\begin{eqnarray}
R_{f}= (1-n_{1f}-n_{4f}+n_{1f} n_{4f}), \quad R_{p}= (1-n_{1p}-n_{4p}+n_{1p} n_{4p}),
\end{eqnarray}
The other $H_0$ eigenvalues and formulae for the effective couplings generated by the RG transformation ($u'_p, u'_f, v', g'_1, g'_2, g'_4, g'_p$ and $g'_n$) are presented in the Appendix.

\begin{figure}
\label{Fig_2}
 \epsfxsize=10cm \epsfbox{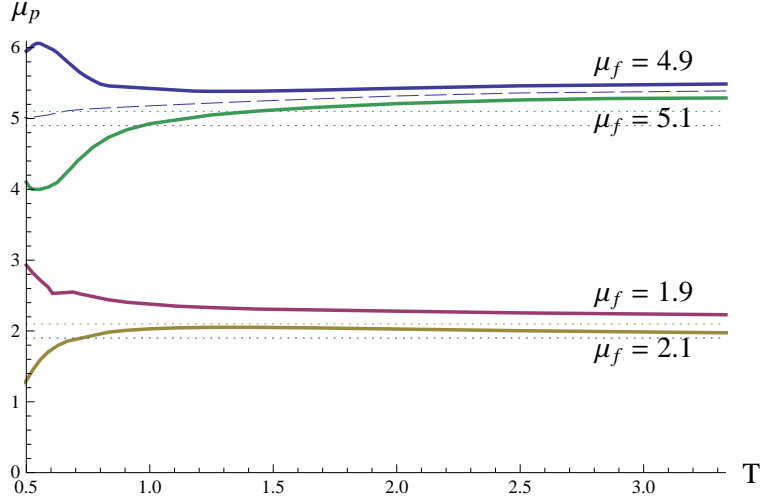}
 \caption{ (Color online) Temperature dependence of the chemical potential $\mu_p$ adjusted to fulfill the condition $<n_f>+<n_p>=1$  for $t_p = 1$, upper curves:  $V = -10$, $\mu_f =4.9, 5.1$ and $5$ (dashed line) and bottom curves: $V=-4$, $\mu_f =1.9,  2.1$.}
\end{figure}

\begin{figure}
\label{Fig_3}
 \epsfxsize=10cm \epsfbox{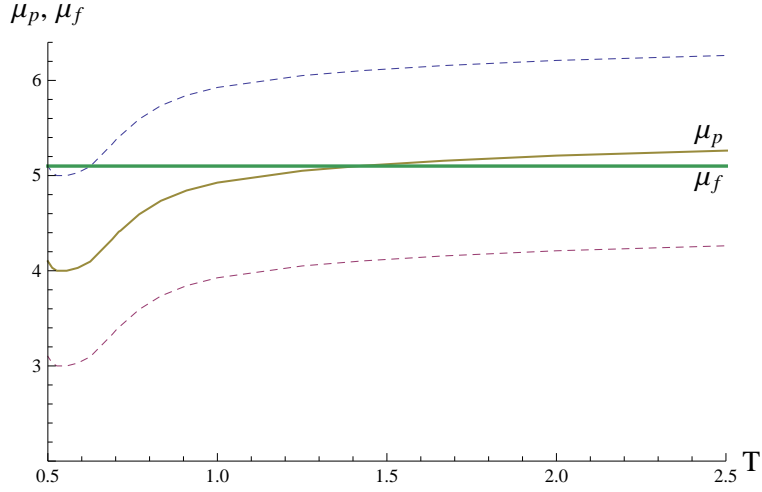}
 \caption{(Color online) The chemical potentials $\mu_f$ and $\mu_p$} for the model with $V=-10, \mu_f=5.1$ and the edges of the $p$-band (dashed lines).
 \end{figure}

     We are now able to evaluate numerically the renormalization transformation (6) (the appropriate recursion relations are given by Eqs (13) and (26)).  The RG transformation allows us to find the thermodynamic properties of the system. The free energy per site can be calculated by using the following formula: 

\begin{eqnarray}
\label{13}
f=\sum_{n=1}^{\infty}\frac{\ln f_0(t_p^{(n)},\mu_p^{(n)},\mu_f^{(n)},V^{(n)}, ...)}{3^n},
\end{eqnarray}
where $"n"$ numbers the RG steps. Knowing the temperature dependence of the free energy one can find the temperature dependences of the internal energy and specific heat but also the average occupation numbers and correlation functions on site $i$ - $G_{p,f}$ and adjacent sites $i,j$ - $G_{p,p} , G_{f,f}$: 
\begin{equation}
G_{p,f} \equiv <n^i_p n^i_f>, \quad G_{p,p} \equiv <n^i_p n^{i+1}_p>,\quad G_{f,f} \equiv <n^i_f n^{i+1}_f>.
\end{equation}

The first step in the LPRG procedure is the choice of the block size. It is obvious that a renormalization group transformation should preserve all symmetries of the original problem. This determines, to some extent, the choice of the block size. For example if one wants to admit the possibility of the existence of a phase transition to the two-sublattice phase one should use blocks with even number of sites $(4, 6, 8,... )$. The advantage from the use of a larger block was discussed in our previous article \cite{SB}. However, in this paper taking into account a number of degrees of freedom for the two-band model, for simplicity, we will confine ourselves to the 4-site block. As shown in Ref.~\onlinecite{SB} for such a block at very low temperature some anomaly in thermodynamic functions is observed, which can be an artefact of neglecting quantum effects between adjacent blocks. So, due to the restricted validity of our procedure at low temperatures all curves are only shown for the reduced temperature $T >0.5$.  

Using the recursion relations (13) and (26) one can calculate the AON for a fixed value of $\mu_f$ and several values of $\mu_p$ and find for a given temperature the value of $\mu_p$ for which the relation $<n_f>+<n_p>=1$ is fulfilled. In Fig.2 the fitted chemical potential $\mu_p$ as a function of temperature is presented for two values of the Coulomb repulsion $V =-4$ with fixed values of $\mu_f =1.9$ and $2.1$ and $V =-10$ with $\mu_f = 4.9, 5.1$ and $5$ (dashed line), all in units of $t_p$. The visible change in the temperature dependence of $\mu_p$ at lower temperature seems to be due to the proximity of  $\mu_f$ to the $p$ - band edge. For example, as seen in Fig.3 for the model with $V=-10$ and $\mu_f=5.1$ the $p$-band edge crosses the level $f$ around $T=0.63$. Evaluating numerically the recursion relation one finds that the RG transformation exhibits only one high-temperature fixed point ($t_p^* =0, V^*$ = const.) as one expects for a one-dimensional system and the system does not undergo any finite temperature phase transition.  Now, using the formula (18) for the free energy per site we can evaluate the average of the band occupation number, specific heat and two point correlation functions. 

\begin{figure}
\label{Fig_4}
 \epsfxsize=16cm \epsfbox{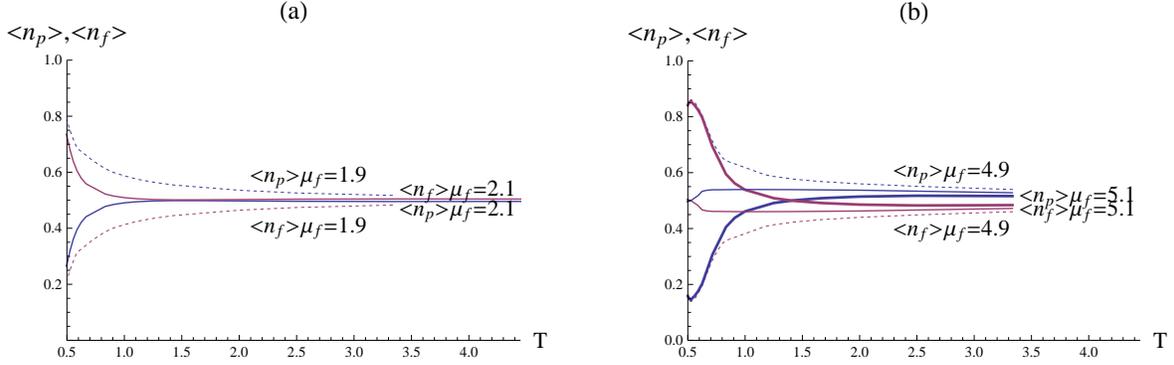}
 \caption{(Color online) Averages of the site occupation as functions of temperature for $t_p=1$, (a) $V=-4$ ($\mu_f = 1.9, 2.1$)  and (b) $V=-10$ and $\mu_f = 4.9, 5.1$ and $5$ (thin lines, upper one  denotes $<n_p>$).}.
 \end{figure}

In Fig.4 the average occupation numbers as functions of temperature are presented for two values of Coulomb repulsion: (i) $V=-4$ with $\mu_f = 1.9$ and $2.1$; (ii) $V=-10$ with $\mu_f = 4.9, 5.1$ and $5$. According to the convention adopted in this paper, where negative $V$ corresponds to the repulsive Coulomb interactions, for $\mu_f > -\frac{1}{2}V$ the $p$ fermions are transferred to the localized state $f$, and vice-versa for $\mu_f < -\frac{1}{2}V$ the localized fermions are transferred to the band with decreasing temperature. As seen in the first case (i) $V=-4$ (left plot)  this transfer is smooth and for $\mu_f >-\frac{1}{2}V$ ($\mu_f < -\frac{1}{2}V$), $n_p > n_f$ ($n_p < n_f$) over the whole range of temperature (from now on we omit the brackets and denote an average as $n_{\alpha}$, $\alpha=f$ or $p$). For larger coupling (ii) $V =-10$ (right plot) we consider three cases. For $\mu_f = -\frac{1}{2}V=5$ the occupation numbers $n_p$ and $n_f$ are almost temperature independent over a wide temperature range. However, due to the hopping term, $n_p > n_f$ for $T > T_{eq} =0.525$. At $T = T_{eq}$ the curves $n_{\alpha}(T)$ intersect and then both AON tend to the value  $n_p = n_f = \frac{1}{2}$. For $\mu_f = 4.9 < -\frac{1}{2}V$ similarly to the weaker coupling case ($V=-4$) $n_p > n_f$ over the whole range of temperature. Differently, for  $\mu_f = 5.1 > -\frac{1}{2}V$ at high temperature $n_p >n_f$, the occupation numbers  are equal to each other at $T_{eq}=1.41$ and then $n_f > n_p$ as expected. 

The specific heat as a function of temperature  for the same models is presented in Fig.5. For the symmetric case $\mu_f= -\frac{1}{2}V=5$ the specific heat curve has a single broad maximum. In other cases the specific heat displays two maxima and a minimum which depth depends on the coupling strength.  A two maxima structure of the specific heat is also observed in the standard one-dimensional Hubbard model with $U > 4t$  and the minimum  corresponds to a maximum in the electronic localization \cite{SB}. In the present model the minima correspond to the maxima in the fermion transfer from the $p$-band to the $f$ - level (for $\mu_f >  -\frac{1}{2}V$) or vice versa from the $f$ - level to $p$ - band (for $\mu_f <  -\frac{1}{2}V$). As seen in Fig.4 indeed, at high temperatures the average occupation numbers  weakly depend on temperature (the transfer between bands is slow) down to a certain temperature at which the transfer rapidly increases. The same two-peak specific heat structure with a sharp peak followed by a broad peak was found for one-dimensional FKM within small cluster exact-diagonalization calculations \cite{PF}. 

Next, we use the RG transformation to find the two-particle on-site $G_{p,f}$ and nearest neighbor (NN) $G_{f,f}$ and $G_{p,p}$ correlation functions (19). In Fig.6 their temperature dependences are shown for $V=-10$ and $\mu_f = 5$ and $5.1$. At high temperatures, all functions tend to $\frac{1}{4}$, which means $\frac{1}{2}$  $p$-particle and  $\frac{1}{2}$ $f$-particle per site, as expected. At low temperature the on-site function $G_{p,f}$  monotonically goes to zero in both cases $\mu_f = 5$ and $5.1$. Instead, the NN function $G_{p,p}$ for the symmetric case $\mu_f = 5$ first slightly increases from $\frac{1}{4}$ ( $G_{f,f}$ decreases) and then tends again to $\frac{1}{4}$. For $\mu_f = 5.1$
$G_{p,p}$ tends to zero.
\begin{figure}
\label{Fig_5}
 \epsfxsize=10cm \epsfbox{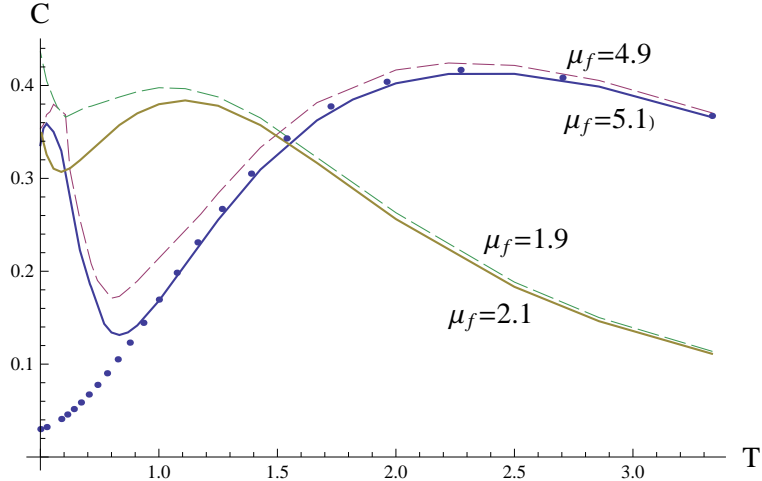}
 \caption{(Color online) Temperature dependence of the one chain specific heat for $t_p=1$, (i) $V=-4, \mu_f=1.9$ and $2.1$ and (ii) $V=-10, \mu_f=4.9, 5.1$ and 5 (dotted line).}
 \end{figure}

\begin{figure}
\label{Fig_6}
 \epsfxsize=10cm \epsfbox{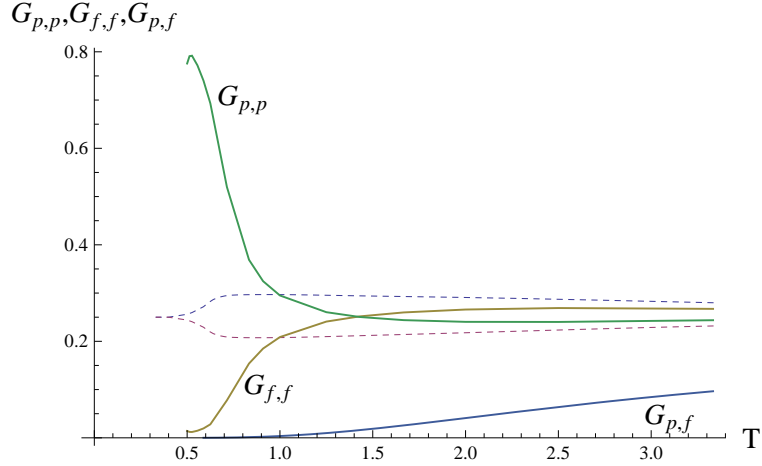}
 \caption{(Color online) Temperature dependence of the correlation functions for $t_p=1, V=-10$ and $\mu_f=5.1$ (solid lines) and    $\mu_f=5.$  (dotted lines, upper one denotes $G_{p,p}$). }
 \end{figure}

\section{Coupled fermion chains}

Below in this section, we shall use the LPRG to study a system with an infinite number of spinless fermion chains at finite temperature, where the chains are coupled by the weak interchain single-particle hopping $t_x$ (3). As was mentioned above, the LPRG transformation when applied on an infinite system generates an infinite number of new interactions already in the lowest nontrivial order of the cumulant expansion. Thus, in order to find the renormalized Hamiltonian we have to confine ourselves to a finite cluster. In a second order calculation one has to consider three rows. We use a cluster with four sites in each row $(4-4-4)$. So the sites from the first and third rows (odd rows) are decimated such that in each RG step every third site survives whereas sites from the second row (even row) are removed (the trace is taken over all sites) \cite{JS}. In addition, for simplification we will consider only two-site interchain interactions thereby neglecting four-site interactions which appear for the $(4-4-4)$ cluster. Under
such an assumption the LPRG transformation generates only one new interaction - an interchain diagonal hopping

\begin{equation}
 t_y \sum_{i,n}( p_{i,n}^\dagger p_{i+1,n+1} +
p_{i+1,n+1}^\dagger p_{i,n}) 
\end{equation}
Hence, in the second order cumulant expansion for the cluster $(4-4-4)$ one has to evaluate the averages (11) of $H_I^2$

\begin{equation}
<H_I^2>_{\alpha} = \big< \left (\sum_{ n=0}^2\sum_{ i=1}^4 \left[ t_x( p_{i,n}^\dagger p_{i,n+1} +
p_{i,n+1}^\dagger p_i) + t_y ( p_{i,n}^\dagger p_{i+1,n+1} +
p_{i+1,n+1}^\dagger p_{i,n})\right ] \right )^2 \big>_{\alpha},
\end{equation}
which means the averages of several fermion operators products e.g:
\begin{equation}
t_x^2 <\sum_{i=1}^4 p^{\dagger}_i P_i P^{\dagger}_{i+1} p_{i+1}>
\end{equation}
where in Eq.(22) the operators $p^{\dagger}, p$ refer to the decimated (odd) rows, and operators denoted by upper-case $P^{\dagger}, P$ refer to the removed (even) rows. These averages have rather complicated expressions and as an example we present the appropriate formula for (22) in the Appendix.

\begin{figure}
\label{Fig_7}
 \epsfxsize=10cm \epsfbox{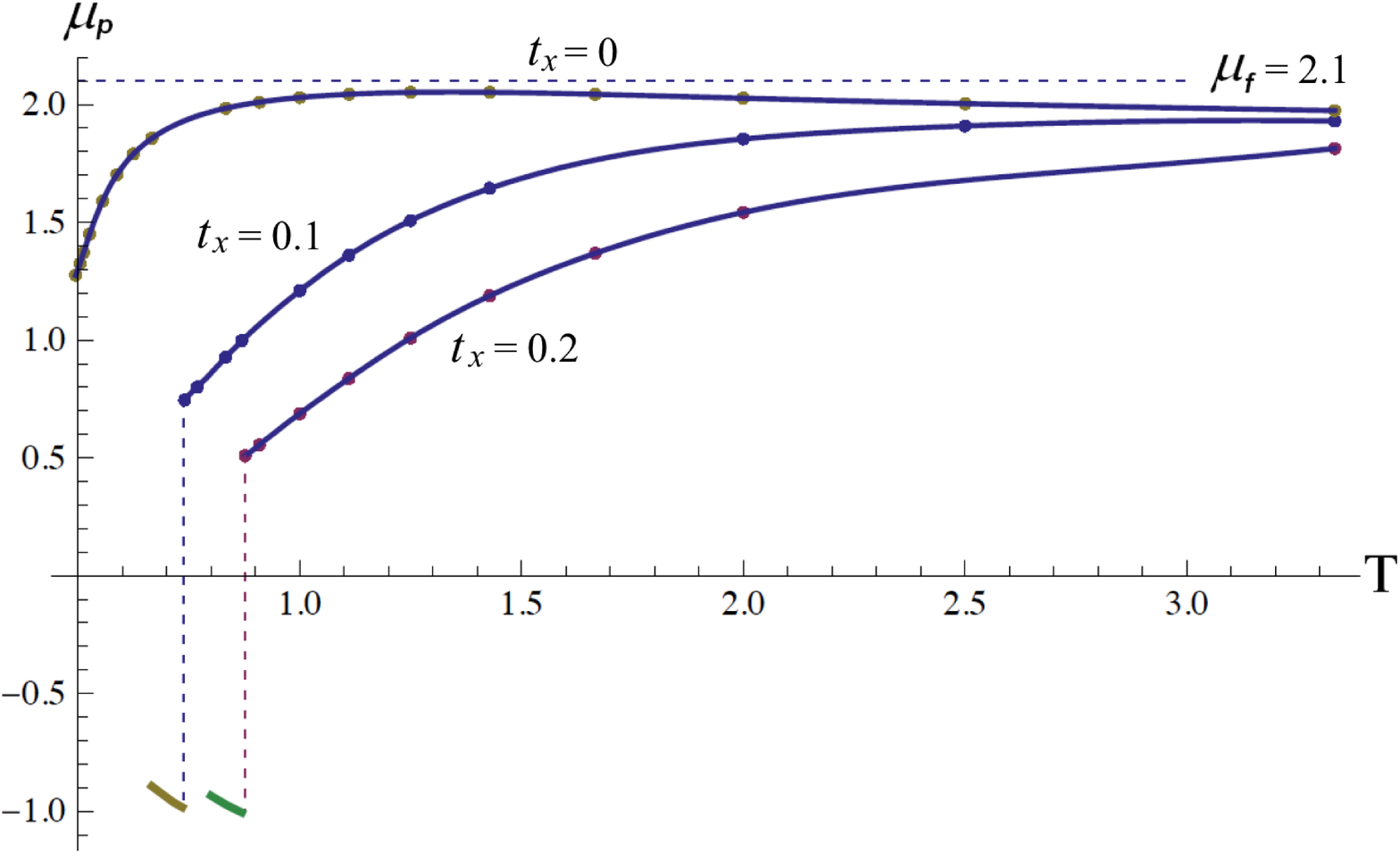}
 \caption{(Color online) Temperature dependence of the weakly coupled chains chemical potential $\mu_p$  for $t_p=1, V=-4$ and $t_x=0, 0.1, 0.2$ (from top to the bottom).}
 \end{figure}

Now, the transformation (10) allows us to find 14 renormalized parameters: four single chain couplings $t_p', \mu'_p,  \mu'_f, V'$ (2), eight created by RG for the single chain (12) and two interchain $t_x', t_y'$ as functions of the original parameters. Knowing the recursion relations for the interaction parameters one can evaluate numerically the LPRG transformation for the Hamiltonian (1) defined by the original parameters $t_p, V, \mu_p, \mu_f$ and $t_x$ and analyze a flow in 14-dimensional coupling parameter space. Now again, we have to find the chemical potential $\mu_p$ for which the condition $<n_p> + <n_f> = 1$ is fulfilled ($\mu_f$ is assumed to be constant). The results are presented in Fig.7 for two values of the interchain coupling $t_x = 0.1$ and $0.2$ and compared with the chemical potential of a single chain. Next we are able to evaluate the specific heat.

\begin{figure}
\label{Fig_8}
 \epsfxsize=15cm \epsfbox{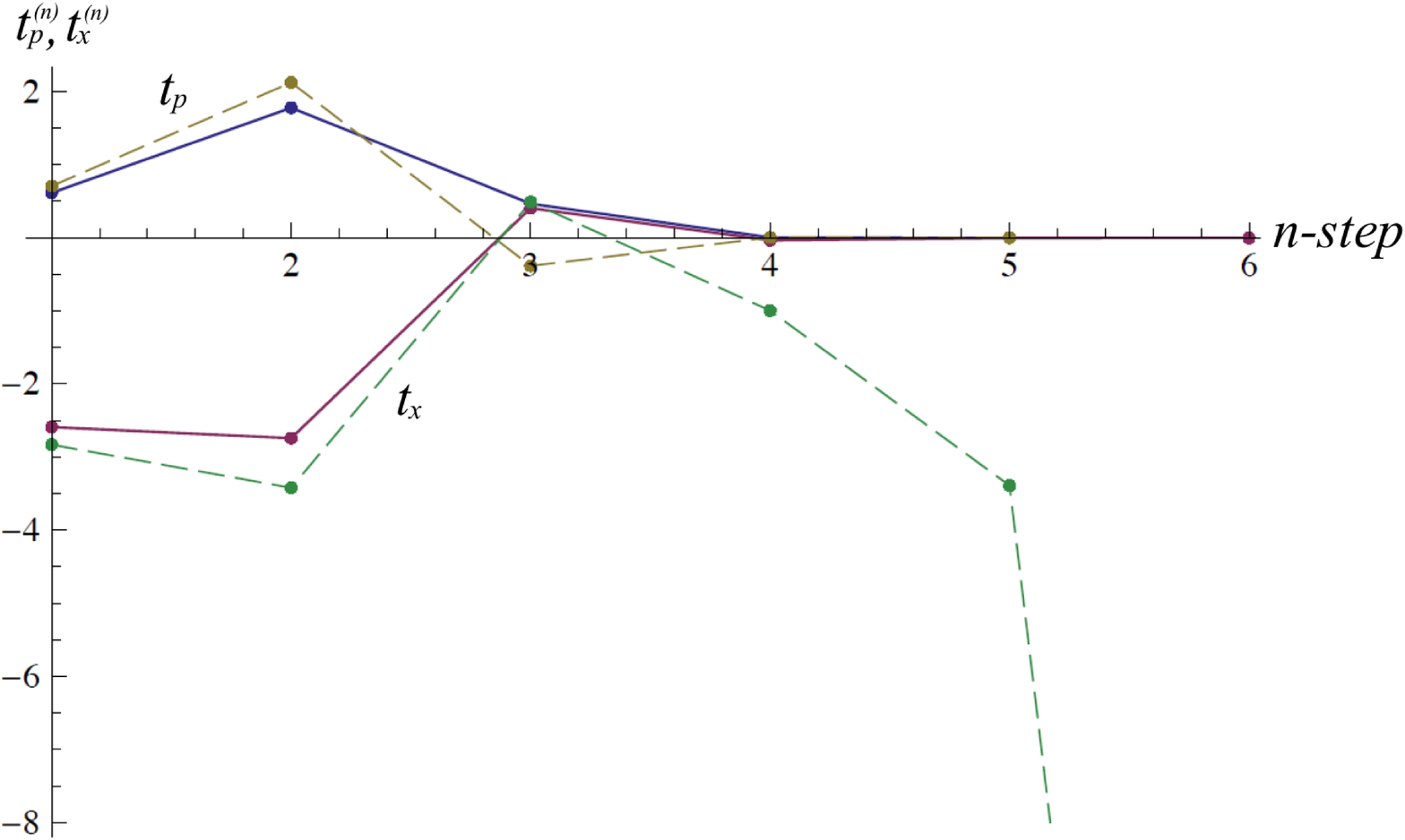}
 \caption{(Color online) The iteration of the parameters $t_p$ and $t_x$ at $T>T^*$ (solid lines) and $T<T^*$ (dashed lines)  for the single-band model (24).} 
 \end{figure}

At this stage it is worthwhile to remind ourselves of the LPRG results for the two-dimensional one-band spinless fermion model at half filling  ($\mu_p = -u_p$) \cite{SB} given by the Hamiltonian

\begin{eqnarray}
\label{1} 
H &=&t_p\sum_{\big < ij\big>}( p_i^\dagger p_j +
p_j^\dagger p_i)+u_p  \sum_i n^i_p n^{i+1}_p +\mu_p \sum_i n^i_p + t_x \sum_{ i,n}( p_{i,n}^\dagger p_{i,n+1} +
p_{i,n+1}^\dagger p_{i,n}) ,
\end{eqnarray}
where $n$ numbers chains.

\begin{figure}
\label{Fig_9}
 \epsfxsize=10cm \epsfbox{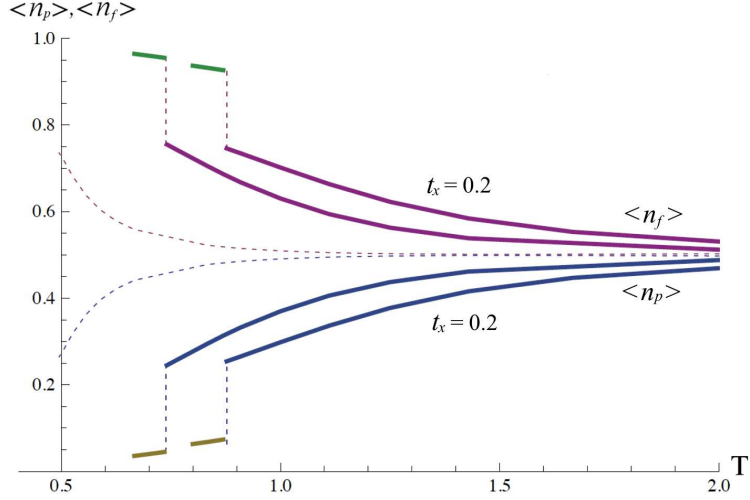}
 \caption{(Color online) Temperature dependences of the coupled chains occupation numbers $<n_f>$ and $<n_p>$ for $t_p=1, V=-4, \mu_f=2.1$ and $t_x=0$ (dashed line) and $t_x=0.1, 0.2$ (solid lines). } 
 \end{figure}

For $t_x \neq 0$ and $T > T^*$ the RG flow  is toward a $T = \infty$ fixed point $t_p^* = 0, t_x^* = 0$ (solid lines in Fig.8)  which describes a disordered phase whereas for $T < T^*$ (dashed lines) the coupling parameter  $t_x$ diverges ($t_x \rightarrow \ -\infty$).
The temperature $T^* = T_c$  can be interpreted as a critical temperature between a disordered phase where the average occupation number of the $p$ fermions is the same for all sites, and a charge ordered phase where this number is different in every other site in a chain. The critical temperature corresponds to the specific heat divergence as seen in Fig.10  (dashed lines). Notice that in the model (24) as well as in the present model there is no Coulomb interaction between the chains which are coupled only by the hopping $t_x$. Thus, the weak interchain hopping in the one band case triggers a charge ordering continuous phase transition. 
\begin{figure}
\label{Fig_10}
 \epsfxsize=10cm \epsfbox{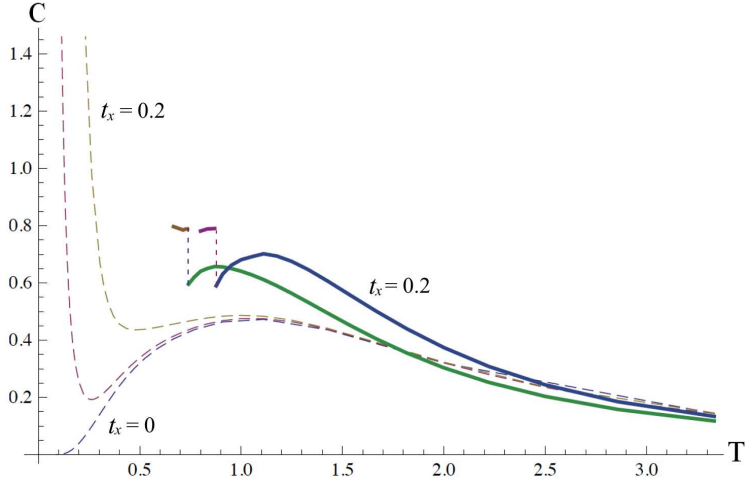}
 \caption{(Color online) Temperature dependence of the coupled chains specific heat for $t_p=1, V=-4, \mu_f=2.1$ and $t_x=0.1,0.2$ (solid lines). For comparison the specific heat for single-band system at half filling with $t_p=1, u_p =-4, \mu_p=4$ for $t_x=0, 0.1$ and $0.2$ is presented (dashed lines). }
 \end{figure}

Let us now turn to the two-band model described by the Hamiltonian (1). As presented in Fig.7, contrary to the one-dimensional case for the coupled chains $\mu_p$ for which the condition (3) is fulfilled, there is a smooth function of T only if the temperature is higher than some $T_b = T(t_x)$ ($T_b \approx 0.74$ for $t_x = 0.1$ and $T_b \approx 0.88$ for $t_x = 0.2$). Technically, if we start with some values of the original parameters for example $\mu_p =2.1, V = -4, t_p = 1, t_x = 0.1$ and iterate the recursion relations for several values of $\mu_p$ at a given temperature, we can find a value of $\mu_p$ that leads to $<n_p>+<n_f> = 1$. It appears that such a continuous solution exists only for $T \ge T_b$. At $T = T_b$ the solution for $\mu_p$ undergoes a jump and consequently the discontinuity in occupation number is observed (Fig.9). This support the claim that the FK model can describe the discontinuous transitions of the $p-$ ($f-$) fermion occupation number as a function of temperature \cite{GF} at least for the weakly coupled chains.The value of the jump decreases upon increasing the interchain coupling $t_x$. Unfortunately, within the present approximation we are not able to decide if the jump  vanishes for the isotropic case ($t_x = t_p$). 

In Fig.10 the specific heat curves of the present model are compared with the results for the weakly coupling chains of the one-band spinless fermion model \cite{SB}. As seen in the latter case the specific heat diverges for a finite value of the coupling parameter $t_x$ as expected at the critical point. For small interchain coupling  ($t_x=0.1$) this divergence is preceded by the hump as a trace of the quasi-one-dimensional character of the system. The hump disappears for larger $t_x$ and for $t_x=0.2$ it is almost invisible. On the contrary, in the present model there is no indication of the continuous phase transition at finite temperature. The specific heat shows a maximum below the temperature at which the band edge ($\mu_p-1$) crosses the level $\mu_f$ (Fig.11) and the discontinuity at $T=T_b$ due to the jump of the occupation number.

\begin{figure}
\label{Fig_11}
 \epsfxsize=15cm \epsfbox{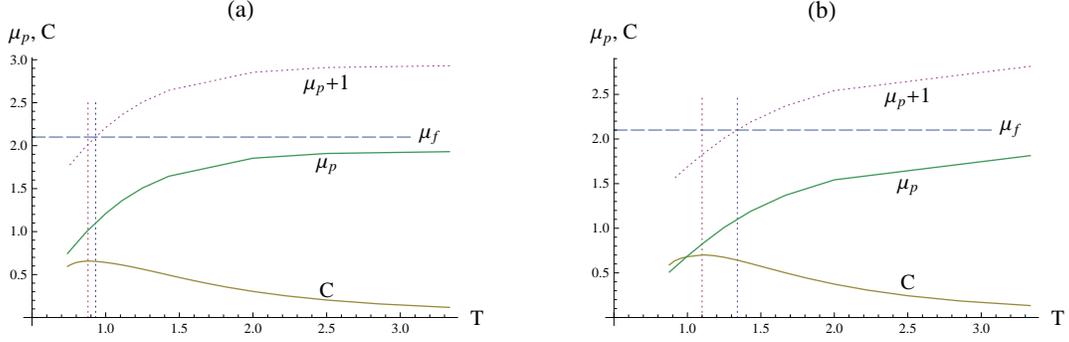}
 \caption{(Color online) Temperature dependence of the coupled chains chemical potentials and specific heat for $t_p=1, V=-4, \mu_f=2.1$ and $t_x=0.1$ (a) and $t_x=0.2$ (b). }
 \end{figure}

\section{Summary}

First, the one-dimensional two-band spinless fermion model with $p$-fermion hopping term $t_p$, on-site  Coulomb repulsion $V$ and chemical potentials $\mu_p,\mu_f$ with one electron per site  (Falicov-Kimball model) has been studied by means of the linear renormalization group transformation. The chemical potential $\mu_p$ has been determined self-consistently by taking into account the conservation of the total number of electrons. The method should lead to reasonable results for $t_p$ not too large (compared with V) and at high temperature. Therefore, two cases have been considered (i) $\mid V \mid = 4 t_p$ and (ii) $\mid V \mid = 10 t_p$ at reduced temperature $T > 0.5$. In both cases the value of $\mu_f$ has been fixed to be slightly below and above $-V/2$. In the first case (i) $\mu_f = 1.9$ and $2.1$ ($V=-4, t_p=1$) and in the second case (ii) $\mu_f = 4.9$ and $5.1$ ($V=-10$). At high temperature the chemical potential is almost temperature independent especially for the weaker coupling ($V=-4$) over a wide temperature interval. At lower temperature, the character of the temperature dependence of $\mu_p$  clearly changes (Fig.2), and the transfer of electrons between the bands rapidly increases (Fig.4). In all cases, the occupation numbers are smooth function of T as expected, only in the case of strong coupling ($V=-10$) and $\mu_f = 5.1 > -\frac{1}{2} V$ the occupation number curves $n_f(T)$ and $n_p(T)$ intersect at $T \approx 1.41$. The specific heat curves exhibit a two-maximum structure found previously for the same model by the small-cluster exact-diagonalization calculations \cite{PF}.

\begin{figure}
\label{Fig_12}
 \epsfxsize=15cm \epsfbox{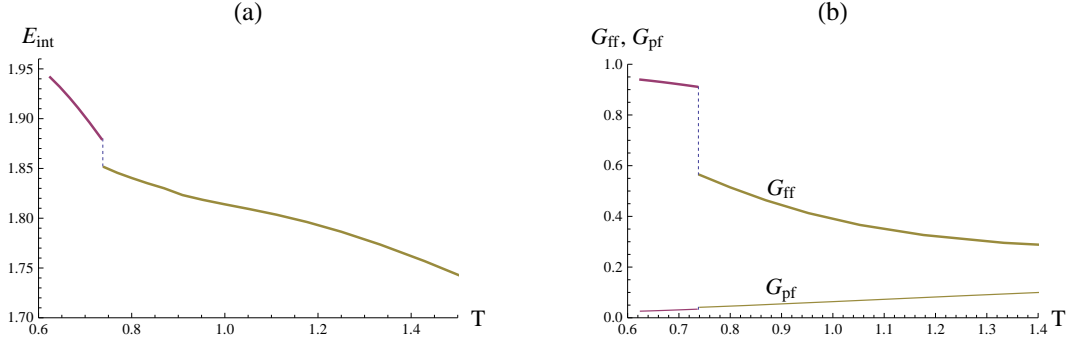}
 \caption{(Color online) Temperature dependences of the coupled chains internal energy (a) and correlation functions $G_{ff}, G_{pf}$ (b) for $t_p=1, V=-4, \mu_f=2.1$ and $t_x=0.1$. }

 \end{figure}
For higher dimension the question of whether the discontinuous transition of the occupation number as a function of temperature occurs in FKM is still an issue of interest. This question was discussed i.a. in the review article of Freericks and Zlati\'c \cite{JF}. The charge-transfer transition in which the character of the electronic states is unchanged, but  their occupancy is shifted from an itinerant to a localized band in FKM was first studied by Falicov, Kimball, and Ramirez \cite{RFK, GF}. Within the molecular field approximation (MFA) they showed that the occupation number $n_f$ for some values of the coupling parameter undergoes a jump at finite temperature. However, in the presumably better approximation (e.g. the coherent-potential approximation), for the same values of the parameters, $n_f$ is smooth function of T \cite{MP,Ghosh}. No discontinuous transition at finite temperature has been found by using small-cluster exact-diagonalization calculations \cite{PF}. Later, Chung and Freericks \cite{CF} showed that for an infinite-coordination Bethe lattice a first-order charge-transfer phase transition can be observed for a narrow value range of the Coulomb interaction. The first order phase transition has also been observed by using the Monte-Carlo method in the weak interaction regime \cite{Maska}.

In this paper to study FKM chains we employed the LPRG method on the $(4-4-4)$-cluster by confining ourselves to two-site interactions only. We have assumed that the chains are coupled by a weak single-particle hopping. Also in this case the chemical potential $\mu_p$ is determined by the condition $n_p+n_f = 1$. However, for the weakly coupled chains the LPRG recursion relations lead to the solution for $\mu_p$ which has a jump. Consequently, there is a discontinuity in the occupation number as a function of temperature (Fig. 9). The value of the jump is smaller for higher interchain hopping and within the present approximation we cannot decide whether a finite jump would also remain for the isotropic system with $t_x = t_p$. The present approach does not permit us to analyze a system below a critical point or spinodal, however we have found discontinuity in the occupation number (Fig.9) and jumps of the internal energy (Fig.12) and the on site $G_{pf} = <n_p^i n_f^i>$ and nearest-neighbor $G_{ff}= <n_f^i n_f^{i+1}>$ correlation functions (Fig.12).  This indicates a discontinuous transition in which the electrons are transferred from the $"p"$ band to the localized $"f"$ level.

Thus, we conclude that the weakly coupled Falicov-Kimball chains with one electron per site ($n_f+n_p =1$) undergoes a finite temperature charge-transfer discontinuous phase transition in which for $\mu_f > -\frac{1}{2}V$ fermions are shifted from an itinerant to a localized band.

\section{APPENDIX}
\begin{appendix}
\subsection{Recursion relations for a single chain}
The eigenvalues of the single chain Hamiltonian (2) completed by the couplings (12) generated by the RG transformation have the form
\begin{eqnarray}
&&\lambda_{10} = f_0+2 f_{\mu_f}+f_{u_f}, \quad \lambda_{(11,12)} = f_0+f_{g_2} \mp f_{g_n} \mp 2 f_{g_p}+2 f_{\mu_f}+f_{\mu_p}  \mp f_{t_p}+f_V+f_v+f_{u_f},
\nonumber\\
&&\lambda_{13} = f_0+2 f_{\mu_p}+f_{u_p}, \quad \lambda_{14} = f_0+2 f_{g_1}+f_{\mu_f}+2 f_{\mu_p}+f_V+f_v+f_{u_p},
\nonumber\\
&&\lambda_{16} = f_0+2 f_{g_1} + 2 f_{g_2} + f_{g_4}+2 f_{\mu_f}+2 f_{\mu_p}  + 2 f_V+2 f_v+f_{u_f}+f_{u_p}.
\end{eqnarray}
where
\begin{eqnarray}
&& f_{v}=Tr_{p,f} (2 n_{1p}-1)(1-n_{4p})(2 n_{4f}-1)(1-n_{1f}) e^{H_0},\nonumber\\
&&f_{u_p}=Tr_{p,f} (2n_{1p} -1) (2 n_{4p}-1) P_f e^{H_0}, \nonumber\\
&&f_{u_f}=Tr_{p,f} (2n_{1f} -1) (2 n_{4f}-1) P_p e^{H_0},\nonumber\\
&&f_{g_1}=Tr_{p,f} (2n_{1p} -1) (2 n_{4p}-1)(1-n_{1f})(2 n_{4f}-1) e^{H_0},\nonumber\\
&&f_{g_2}=Tr_{p,f} (2n_{1p} -1)(1-n_{4p})(2 n_{1f}-1)(2 n_{4f}-1) e^{H_0},\nonumber\\
&&f_{g_4}=Tr_{p,f} (2n_{1p} -1)(2 n_{4p}-1)(2 n_{1f}-1)(2 n_{4f}-1)  e^{H_0},\nonumber\\
&&f_{g_p}=Tr_{p,f} p_1 p_4^\dagger(2 n_{1f}-1)(1-n_{4f}) e^{H_0},\nonumber\\
&&f_{g_n}=Tr_{p,f} p_1 p_4^\dagger(2 n_{1f}-1)(2 n_{4f}-1) e^{H_0}.
\end{eqnarray}
and renormalized couplings are
\begin{eqnarray}
&&v' =\frac{1}{2} \lg \frac{f_0^2}{\lambda_2^2 \lambda_4 \lambda_5}+\frac{f_V-f_v}{2 Q} \lg \frac {\lambda_6}{\lambda_8}+\frac{1}{2}\lg (\lambda_6 \lambda_8), \quad
u'_p =\lg \frac {f_0 \lambda_{13}}{\lambda_4 \lambda_5},\quad 
u'_f =\lg \frac {f_0 \lambda_{10}}{\lambda_{12}^2},  \nonumber\\
&& g' _ 1 =\lg \frac {\lambda_4 \lambda_5 \lambda_{12} \lambda_{14}}{f_0 \lambda_{13}\lambda_{16} \lambda_{18}}, \quad
g' _ 2 =\frac{1}{2} \lg \frac{\lambda_2^4 \lambda_4 \lambda_5 \lambda_8 \lambda_{11}}{f_0^2 \lambda_{10}^2 \lambda_6^2 \lambda_8^2} - R \lg (\lambda_{12}),\quad  \nonumber\\
&& g' _ 4 =\lg \frac{f_0 \lambda_6^2 \lambda_8^2 \lambda_{10} \lambda_{12} \lambda_{13} \lambda_{16}}{\lambda_2^2 \lambda_4 \lambda_5 \lambda_{11} \lambda_{14}^2}+R \lg (\lambda_{12}), \quad
 g' _p =\frac{1}{2} \lg \frac{\lambda_4}{\lambda_5}-\frac{f_{g_p}+f_{t_p}}{W} \lg\frac{\lambda_6}{\lambda_8},\quad  \nonumber\\
&& g' _n =\frac{1}{2} \lg \frac {\lambda_5}{\lambda_4 \lambda_{11} } +
\frac {2 (f_{g_p}+f_{t_p})} {W} \lg \frac {\lambda_6}{\lambda_8} -\frac{1}{2} R \lg (\lambda_{12}), \quad
R = \frac {f_{g_n}+2 f_{g_p}+f_{t_p}}     {f_{g_n}-2 f_{g_p}-f_{t_p}}.
\end{eqnarray}

\subsection{Coupled chains}

To evaluate the transformation (10) one has to know the averages of the products of the original fermion operators from the decimated ($p$) and removed ($P$) rows of a type $<p^{\dagger}_i P_k P^{\dagger}_l p_j>$. All of them are expressed through the effective fermion operators ($\sigma^{\dagger}, \sigma, \phi^{\dagger}, \phi$) and, for example,

\begin{eqnarray}
&&<\sum_{i=1}^4 p^{\dagger}_i P_i P^{\dagger}_{i+1} p_{i+1}> = -\sum_{i=1}^4<p^{\dagger}_i p_{i+1}><P^{\dagger}_{i+1} P_i>  = -r1 <\hat Q>\nonumber \\
&& = -r_1 (Q_{10}+Q_{\mu_p}(n^{(1)}_{\sigma}+n^{(2)}_{\sigma})+Q_{\mu_f}(n^{(1)}_{\phi}+n^{(2)}_{\phi})+Q_{u_p}(n^{(1)}_{\sigma} n^{(2)}_{\sigma})+Q_{u_f}(n^{(1)}_{\phi} n^{(2)}_{\phi})\nonumber \\
&&+Q_{V}(n^{(1)}_{\sigma} n^{(1)}_{\phi}+n^{(2)}_{\sigma} n^{(2)}_{\phi})+Q_{v}(n^{(1)}_{\sigma} n^{(2)}_{\phi}+n^{(2)}_{\sigma} n^{(1)}_{\phi})+Q_{t_p}(\sigma_1^{\dagger}\sigma_2+\sigma_2^{\dagger}\sigma_1).
\end{eqnarray}
where
\begin{equation}
\hat Q=\sum_{i=1}^4 p^{\dagger}_i p_{i+1},
\end{equation}
and
\begin{eqnarray}
&&Q_{10}=C_0 Z_0, \quad Q_{\mu_k}=C_{\mu_k} Z_0+C_0 Z{\mu_k}+C_{\mu_k} Z_{\mu_k}, \quad (k=p,f) \nonumber\\
&&Q_{u_k}=C_0 Z_{u_k}+2 C_{\mu_k}(Z_{\mu_k}+Z_{u_k})+C_{u_k}(Z_0+2Z_{\mu_k}+Z_{u_k}), \nonumber\\
&&Q_{V}=C_{\mu_f} Z_{\mu_p}+ (C_0+C_{\mu_f}) Z_V
+C_{\mu_p}  (Z_{\mu_f}+Z_V)+C_V(Z_0+Z_{\mu_f}+Z_{\mu_p}+Z_V), \nonumber\\
&&Q_v=C_{\mu_p}Z_{\mu_f}+C_{\mu_f}Z_{\mu_p}+C_v(Z_0+Z_{\mu_p}+Z_{\mu_f}),
\end{eqnarray}
with
\begin{eqnarray}
&&Z_0=Tr_{p,f} \hat {Q} R_f R_p e^{H_0},\quad Z_{\mu_k}=Tr_{p,f} \hat {Q}(2n_{1k}-1)(1-n_{4k})R_k e^{H_0}, \nonumber\\
&&Z_{u_k}=Tr_{p,f} \hat {Q}(2n_{1k}-1)(2n_{4k}-1)R_{k'} e^{H_0},\nonumber\\
&& Z_{V,v} = Tr_{p,f} \hat {Q} (1-2 n_{1p}) (1-n_{4p})(1-n_{(4,1)f}) e^{H_0},\nonumber\\
&&Z_{t_p}=Tr_{p,f} \hat {Q} p_1 p_4^{\dagger} e^{H_0},  \qquad (k,k' = p,f), \nonumber\\
&&C_0 = \frac{1}{f_0},\quad C_{\mu_p}=\frac{1}{\lambda_4}+\frac{1}{\lambda_5}-\frac{2}{f_0},\quad C_{\mu_f}=2(\frac{1}{\lambda_2}-\frac{1}{f_0}),\quad C_{t_p}=\frac{1}{2}(\frac{1}{\lambda_5}-\frac{1}{\lambda_4}), \nonumber\\
&&C_V=\frac{2}{f_0}-\frac{2}{\lambda_2}-\frac{1}{\lambda_4}-\frac{1}{\lambda_5}+\frac{1}{\lambda_6}+\frac{1}{\lambda_8}+\frac{f_v-f_V}{\lambda_6 \lambda_8}, \nonumber\\
&&C_v=\frac{1}{f_0}-\frac{1}{\lambda_2}-\frac{\lambda_4+\lambda_5}{2 \lambda_4 \lambda_5}+\frac{f_v-f_V+\lambda_6+\lambda_8}{2 \lambda_6 \lambda_8}. 
\end{eqnarray}

For the decimated rows the single chain averages are given by
\begin{equation}
<P_1^{\dagger}P_{1+\alpha} > = \frac{Tr[ P_1^{\dagger}P_{1+\alpha} e^{H_0}]} {Tr[e^{H_0}]}.
\end{equation}

\end{appendix}


\end{document}